# Enabling Automatic Disordered Speech Recognition: An Impaired Speech Dataset in the Akan Language

*Isaac Wiafe, Akon Obu Ekpezu, Sumaya Ahmed Salihs, Elikem Doe Atsakpo, Fiifi Baffoe Payin Winful, Jamal-Deen Abdulai*

*Department of Computer Science, University of Ghana*

**Abstract**

The lack of impaired speech data hinders advancements in the development of inclusive speech technologies, particularly in low-resource languages such as Akan. To address this gap, this study presents a curated corpus of speech samples from native Akan speakers with speech impairment. The dataset comprises of 50.01 hours of audio recordings cutting across four classes of impaired speech namely stammering, cerebral palsy, cleft palate, and stroke induced speech disorder. Recordings were done in controlled supervised environments were participants described pre-selected images in their own words. The resulting dataset is a collection of audio recordings, transcriptions, and associated metadata on speaker demographics, class of impairment, recording environment and device. The dataset is intended to support research in low-resource automatic disordered speech recognition systems and assistive speech technology.

## 1 Background

Speech impairment affects communication, education, and social participation. However, most existing speech datasets and assistive speech technologies, such as Alexa and Siri, focus on clinically standardized speech. These systems are not designed to accurately recognize the speech patterns of users with impairments. The challenge is further exacerbated by the technological divide between high-resource and low-resource languages. Akan, a major indigenous language spoken in Ghana, remains particularly underrepresented in impaired speech corpora. As of the time of this study, no publicly available data or studies exist on automatic disordered speech recognition (ADSR) in Akan. Previous attempts to digitalize the Akan language have predominantly focused on standard speech [1–3,5].

To address this gap, this study presents a dataset of impaired speech in Akan. The dataset comprises approximately 50-hours of audio recordings in Akan from four aetiologies of impairment namely cleft palate, cerebral palsy, stammering, and stroke-induced speech disorder. This study therefore provides a foundation for inclusive speech modeling and the development of technologies capable of recognizing impaired speech in low-resource language contexts. The potential of part of this dataset in ADSR has been demonstrated in Salihs et al [4]. The remaining sections of this paper refer to the dataset as UGAkan-ImpairedSpeechData [6].

Correspondence: aoekpezu@st.ug.edu.gh

## 2 Methods and Materials

### 2.1 Materials

This study utilized two main tools for data collection and transcription: the UGSpeechData mobile app and a custom desktop transcription interface. Each tool is described in detail in the subsections that follow.

#### 2.1.1 The UGSpeechData Mobile App

This study utilised the UGSpeechData mobile application, which was originally designed and previously employed to collect standard speech data for five Ghanaian languages [5]. The app is an Android-based tool designed to record audio, track recording sessions, validate, transcribe and export transcriptions to cloud servers for secure storage. To ensure that the UGSpeechData mobile app could effectively support the collection of impaired-speech data, several modifications were implemented to address the unique speech patterns and recording needs of individuals with various aetiologies of impairment. These adaptations were essential to preserve the authenticity of impaired speech, enhance the descriptive prompts used during recordings, and enable accurate categorisation of participants. Table 1 summarises the key differences between the original version of the app used for collecting standard speech and the modified version employed in this study.

Table 1: Modifications to the UGSpeechData App for Impaired-Speech Data Collection

| Component | Standard Speech Version | Impaired Speech Version | Rationale |
|---|---|---|---|
| Pause / Break Handling | Restricted frequent pauses; recordings with more than 3 seconds of silence or excessive filler sounds were not saved. | Removed all pause restrictions, allowing recordings with natural pauses, fillers, disfluencies, and slurred segments. | Impaired speech often includes frequent pauses, disfluencies, and slurring as part of natural speech patterns. Restricting these behaviours would have resulted in the loss of authentic speech patterns. |
| Prompt Images | Included 1,000 images for eliciting spontaneous descriptions. | Added 200 new images (total = 1,200), including simpler and more easily identifiable images. | Enlarging the image set especially with simpler and more recognisable images increased vocabulary coverage and elicited easier descriptions or storytelling, particularly for individuals with impaired speech who may also have some level of cognitive disability. |
| Maximum Recording Duration | Allowed a 15–30 second limit for each audio file. | Extended duration to up to 60 seconds per recording. | Impaired speech often required more time due to slow articulation, repetitions, hesitations, and pauses. A longer duration ensured that complete utterances were captured without interruption. |
| User Profile Fields | Collected basic demographic information only. | Added aetiology information to classify each user's type of speech impairment. | Aetiology tagging enabled accurate categorisation and strengthened the dataset structure. |
| Validation/Transcription | Validation and transcription were performed within the mobile app using built-in features on Android devices. | These features were disabled. | To enhance efficiency, support longer audio recordings, provide a larger interface, and enable accurate rendering of Akan-specific characters using a dedicated Akan keyboard. |

Correspondence: aoekpezu@st.ug.edu.gh

*2.1.2 Transcription App*

As shown in Table 1, the transcription workflow used for the impaired-speech dataset differed significantly from the workflow used for the standard-speech collection. While the UGSpeechData mobile app originally supported in-app validation and transcription on Android devices, these features were disabled for the impaired-speech project. The decision to separate transcription from the mobile app was driven by the needs of the language experts who performed the transcriptions. They required a desktop-based tool that offered a larger interface, more comfortable navigation, and support for extended transcription sessions involving longer and more complex speech recordings.

To meet these requirements, a custom desktop transcription app was developed for Windows (https://github.com/HCI-LAB-UGSPEECHDATA/Transcription-App). The transcription tool (see Figure 2) was built in Python, using Tkinter for the graphical interface, Pygame for audio playback, and Pillow for image rendering. It loads each audio file alongside its corresponding image prompt, thus displaying an intuitive interface that enabled transcribers to listen to the audio and input transcriptions in real time.

To ensure accurate linguistic representation, the transcription app was complemented by a custom Akan keyboard used in [5]. This keyboard (https://github.com/HCI-LAB-UGSPEECHDATA/SPEECH-DATA-KEYBOARDS/blob/main/akan.kmp) supports the accurate input of Akan-specific characters such as "ɔ" which are often misrepresented with symbols such as the close bracket ")" in the standard system keyboards.

The transcription app provided standard playback controls (play, pause, resume, replay) and includes a visual progress bar, a transcription status indicator, and "Next"/"Previous" navigation buttons to move efficiently through the dataset. Transcriptions were entered into an integrated text field and automatically stored in an editable Excel spreadsheet, which updates continuously to prevent data loss. The tool was packaged with PyInstaller to facilitate easy distribution and installation across multiple computers.

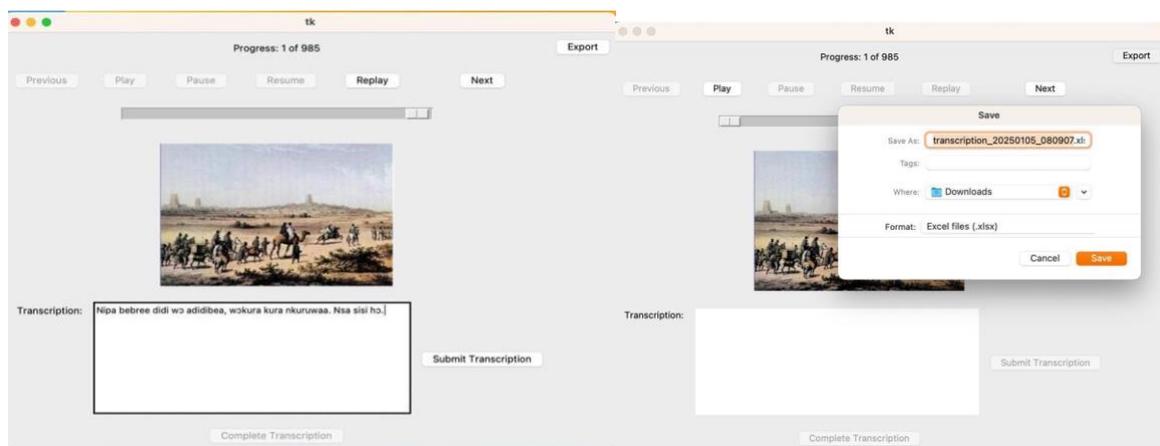

Figure 1. Transcription App interface (left: app in use), right (export)

## 2.2 Methods

This study followed a multi-stage methodological approach encompassing participant screening and recruitment, audio recording using the modified UGSpeechData mobile application, validation of collected recordings, and transcription using a custom desktop interface. The procedures were designed to ensure the reliable collection of impaired-speech data while maintaining ethical standards, linguistic accuracy, and methodological rigor. The subsections that follow describe each stage in detail.

*2.2.1 Participant screening, selection, and informed consent*

Participants were recruited through convenience sampling. Speech therapists based in Accra and Kumasi assisted in identifying individuals with cerebral palsy and stammering, primarily from institutions in Kumasi. Recruiting from Kumasi was essential to ensure that the recordings reflected the speech patterns of native Asante speakers. This was because the project primarily focused on the Asante dialect of the Akan language. Thereafter, to capture broader linguistic variation within Akan, additional participants were recruited from Accra. This group included individuals recovering from stroke and those with cerebral palsy, cleft palate, and stammering.

User accounts were created by members of the research team, with each account assigned an app-specific email in the format akanXX@ugspeechdata.com, where XX denoted the unique number allocated to each participant for identification and data management. Participants accessed the mobile app using the provided email and password to sign in and record audio. Users were encouraged and when necessary, assisted to confirm and update their profiles (names, age, gender, and class of aetiology) immediately after logging in. Within the profile section, participants were required to read and accept the informed consent.

Informed consent was obtained directly through the app. The consent form clearly communicated the purpose of the project, the intended use of the collected data, participants' right to withdraw at any time, and the details of compensation. Refer to Figure 3 for screenshot of the consent page. Additionally, all participants were required to confirm that they were 18 years or older.

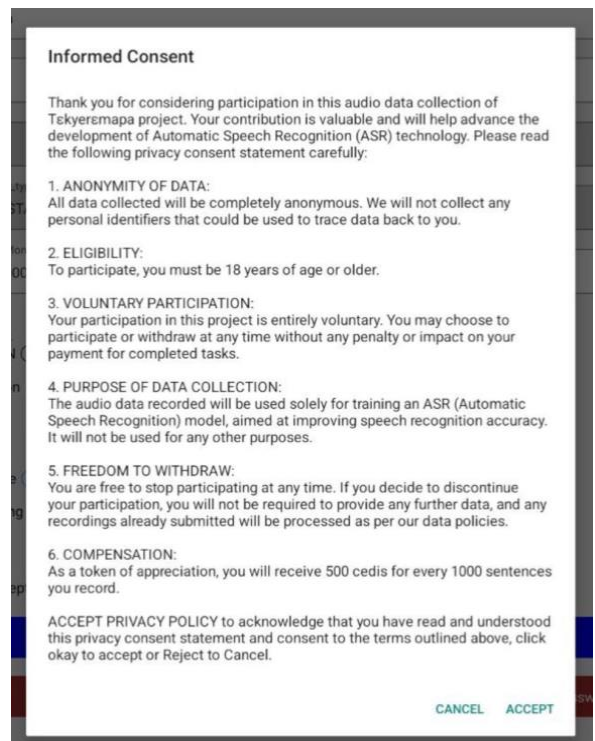

Figure 2. Screenshot of informed consent

Correspondence: aoekpezu@st.ug.edu.gh

### 2.2.2 Audio Recording

To elicit spontaneous natural speech, participants were asked to describe a set of provided images in their own words. The image-prompt approach was adopted because it has been shown to yield natural speech recordings and generate varied expressions based on the same visual stimulus image [5]. The mobile app also allowed users to track the number of audio clips they had recorded and uploaded.

Due to cognitive and mobility challenges particularly among participants with cerebral palsy some individuals required assistance during the recording process. Speech and language therapists, together with members of the research team, supported these participants by helping them hold the mobile device and by facilitating the speech elicitation process. This sometimes involved telling brief stories or providing contextual cues about the picture to help participants generate descriptions before recording.

The following recording guidelines were provided to participants and were subsequently applied by validators and transcribers to flag non-conforming audio files:

i. Recordings should be made in a quiet environment, free from background noise.
ii. All recordings must be in the Akan language; English was permitted only when no equivalent Akan term existed.
iii. Each recording should be between 5 and 60 seconds in duration.
iv. Participants were encouraged to speak freely and naturally, especially individuals who stammer, to discourage self-censorship.
v. Recordings should not begin with filler phrases such as "I see," "in this picture," or "it looks like."
vi. Recordings must not contain profane or offensive language.

The recording feature within the app was only activated when the environmental noise level was sufficiently low; otherwise, the app displayed a "too noisy" notification and prevented recording. Participants could skip any prompt they were unable to describe. After each session, the app allowed users to play back, save, or delete the recording, and they could exit the app at any time to take breaks.

### 2.2.3 Audio Validation and Transcription

Following the collection of audio recordings, validation and transcription were carried out by eight linguistic experts, all indigenous speakers of Asante Twi with university-level training in Akan. Prior to the main exercise, the experts received training on how to use the desktop transcription application, how to export transcriptions, and how to apply both the validation and transcription guidelines.

Validation and transcription were conducted simultaneously in a structured two-phase workflow designed to enhance accuracy, reliability, and quality control. The eight experts were blindly paired into two groups, with each group assigned the same set of audio files without their awareness of this duplication. This design introduced an element of inter-rater reliability and allowed the research team to detect discrepancies, verify agreement between independent transcribers, and minimise subjective bias in interpreting impaired speech.

The audio recordings were divided into batches of approximately 500 files per batch, grouped as much as possible by speaker. This approach proved advantageous because impaired-speech characteristics vary significantly across individuals; thus, increased familiarity with a speaker's

Correspondence: aoekpezu@st.ug.edu.gh

patterns helped experts better understand, interpret, and accurately transcribe subsequent recordings.

Each linguistic expert received and processed audio files in successive batches of 500, completing both validation and transcription for a batch before being assigned the next one. This iterative process continued until all 14,312 audio recordings had been processed by the entire team. On average, each expert validated and transcribed approximately 1,800 audio files, though the exact number varied slightly depending on the rate at which individual experts completed earlier batches.

In the second phase referred to as transcription validation, the research team compared the two independent transcripts generated for each audio file. A transcript was accepted if both linguists produced identical or near-identical outputs. For files with conflicting transcriptions, the audio was reassigned to a new pair of experts who had not previously worked on that file for independent review and conflict resolution. This multilevel approach strengthened dataset reliability, minimized transcription errors, and ensured consistency across the entire corpus.

For audio validation, linguists were required to identify and flag unacceptable content, including offensive or profane language, hate or abusive expressions, and personal identifiers (PI), such as names or phone numbers. These were flagged by typing the keyword "reject" in the transcription field.

Given the complexity of impaired speech and the need for consistency across multiple transcribers, it was essential to establish clear and detailed transcription rules. These guidelines which were co-designed with the linguistic experts governed the handling of repetitions, disfluencies, speech errors, adapted English words, ambiguous utterances, and other transcription challenges.

General guidelines were agreed to handle non-verbal sounds such as "εεεεεεε, εε, εεε" as "εεε", and "aaaaaaa" as "aaa", also "mmmmm" as "mmm". Certain words in Akan may have alternative spellings while conveying the same meaning. For example, Tɛkyerɛma is equivalent to Tɛkerɛma, soronko is equivalent to sononko, and nneɛma is equivalent to nnooma. However, it is important to note that the spoken form of a word may differ from its standard written version. For instance, Ɔmo is spelled as wɔn in standard written Akan. Transcribers are expected to adhere to the correct conventions. The main guidelines are presented in Table 2.

Table 2: Transcription guidelines

| Acceptable Practices | Unacceptable Practices |
| --- | --- |
| Transcribe exactly what you hear as spoken, ensuring that all utterances from the primary speaker are included and that the original spelling of each word is maintained. | Do not correct any grammatical errors and avoid transcribing audio that contains background noise or overlapping speech from a secondary speaker. |
| Include self-repetitions; if a speaker repeats a word, phrase, or sentence, transcribe it exactly as spoken. | Do not omit repetitions, even if they appear unnecessary. |
| Capture self-corrections by transcribing everything, including when a speaker changes a word or phrase or corrects themselves mid-sentence. | Do not omit self-corrections; both attempts at a word or phrase should be retained. Do not modify the word order; preserve the natural flow of speech. |
| Listen to 3 to 4 words at a time and transcribe in small segments to ensure accuracy. | Do not attempt to memorize what you hear and recall it later while typing. |
| Use standard Akan punctuation and capitalization rules. For e.g., write Tɛkyerɛma instead of T3kyer3ma, or Nnooma instead of Nnocma. | Do not alter punctuation, capitalization, or characters in a way that goes against standard Akan rules. |

Correspondence: aoekpezu@st.ug.edu.gh

| | |
|---|---|
| Spell out numbers and flag phone numbers, IDs as PII (Personally Identifiable Information). | Do not use numerals, and do not transcribe phone numbers or IDs. |
| Transliterate English words in Asante language-specific form (e.g., write "kɔmputa" instead of "computer" even if the speaker said Computer). | Do not automatically write English words if you do not know its Asante language-specific form. Flag it with the keyword "language". This will be addressed in the transcription validation phase. |

Impaired speech often exhibits distinctive patterns such as stretched words, split words, un-ended words, repetitions associated with standard disfluencies, and stammering-related blocking and repetitions. To ensure these patterns were handled consistently across the dataset, the transcribers and the research team engaged in a series of deliberations to reach a consensus on the appropriate transcription approach for each pattern. This collaborative process resulted in a set of impaired-speech–specific guidelines that promoted uniformity and accuracy throughout the transcriptions. Table 3 outlines these speech patterns and the corresponding recommended actions.

Table 3: Impaired speech specific patterns and their transcription guidelines

| Specific patterns | Recommended Actions |
|---|---|
| Stretched words | To capture the full acoustic form of the utterance, try as much as possible to preserve the elongated sound pattern. Thus, transcribe the word exactly as it is stretched and pronounced. For example, if the speaker says "baaaanku", the transcription should retain the stretched form (e.g., "baaaanku") rather than shortening it to "banku". |
| Split words | Transcribe the word or phrase exactly as it is produced by the speaker. For instance, "Anidie" (meaning respect) is the correct and intended pronunciation and should be transcribed as "Anidie" when spoken as a single unit. However, if the speaker due to their speech pattern splits the word into two parts, such as "Ani die" (which literally translates to "Eye eat"), then the transcription should retain this split form. This ensures that the transcript accurately reflects the speaker's articulation rather than the linguistically correct form. |
| Un-ended words (incomplete articulations) | Transcribe the complete word as intended by the speaker, using contextual cues from the audio or the associated image. For example, if the speaker says "ased" but the context of the description indicates that they intended to say "aseda", the transcription should reflect the full intended word ("aseda"). |
| Repetitions for standard disfluencies | Transcribe all repetitions exactly as spoken, ensuring that each repeated word, phrase, or segment is fully reflected in the transcript. For example, if the speaker says, "I want I want banku", the transcription should preserve the repetition as "I want I want banku." This approach was expected to ensure that natural disfluencies are accurately represented and not simplified or omitted. |
| Disfluency based on stammering and blocking | Transcribe all repetitions exactly as spoken, preserving every instance of stammering or blocking. For example, if the speaker says "Yɛ yɛ yɛ yɛfrɛ", the transcription should retain the full sequence exactly as produced. This approach ensures that all disfluent patterns are accurately captured, without omission or smoothing, thereby reflecting the true acoustic characteristics of the speaker's impaired speech. |

Given the inherent difficulty of deciphering some impaired-speech recordings, the experts and the research team recognised the need to establish clear and consistent guidelines for handling uncertainties. These discussions resulted in an agreed framework outlining scenarios in which transcribers might be unsure about the content of an audio file and the appropriate actions to take in such cases. The purpose of these guidelines was to ensure that only recordings the transcribers felt confident about were included in the dataset, thereby preventing poor-quality or inaccurate transcriptions. This systematic approach also enabled efficient identification and filtering of files flagged with issues, which were subsequently excluded during the post-transcription phase. Table 4 presents the key uncertainty scenarios and the corresponding recommended actions and flagging conventions.

Correspondence: aoekpezu@st.ug.edu.gh

Table 4: Guidelines for handling uncertainties and issues in transcription

| Scenario | Action | Flagging Approach **(type the keyword "issue: describe the issue in your own words as much as possible"** |
|---|---|---|
| Inaudible Audio or Unclear Speech | Listen to the segment multiple times, adjusting the speed. If still unclear, flag the audio. | "Issue: Inaudible" or "Issue: Incoherent" |
| Audio in a Language Outside Assigned Scope | If the audio is in a language other than Akan, do not attempt transcription. | If the other language is known, e.g., Ga, then type "Issue: Language [Ga]" or "Issue: Language[N/A]" (if language is unknown) |
| Uncertainty About Transcription Accuracy | Mark the level of uncertainty regarding the transcription. | "Issue: Relatively certain, but with some doubt" or "Issue: Very uncertain or Issue: Could not understand" |
| Profanity or Offensive Language | Do not transcribe audio containing profane, vulgar, or hate speech. | "Issue: Profane/Vulgar" or "Issue: Hate Speech/Offensive Language" |
| Presence of Phone Numbers or Personal Identifiers (PIIDs) | Flag phone numbers and other personal identifiers in the audio. | "Issue: PII" |

## 3 Data Description

### 3.1 Summary of the Dataset

The UGAkan-ImpairedSpeechData is a dataset that comprises of 14,312 audio files and corresponding transcriptions amounting to 50.01 hours. The audio files are a collection of spontaneous image descriptions from four distinct aetiologies of impairment, namely stammering, cerebral palsy, cleft-plate, and stroke-induced speech disorder, each exhibiting unique acoustic and linguistic features. Recordings were done in different environments, including outdoor (7,706 audio files), unspecified (3,075), indoor (2,254), studio (982), and car (295).

The UGAkan-ImpairedSpeechData is organised into five subfolders: four folders representing the individual aetiologies and an images folder containing the prompt used for eliciting descriptions. Each aetiology folder includes:

  i. an "audios" folder, which holds the MP3 audio files and
  ii. a metadata CSV file, named using the format *metadata_[aetiology].csv* (e.g., *metadata_stroke.csv* for the stroke category), which provides structured information about each audio file.

Each metadata file contains thirteen attributes describing the speaker, recording context, and audio characteristics. These include the original file name (orig_file_name), impairment type (condition), gender, speaker identifier (speaker_id), image prompt (images), recording environment, age, device used, transcription text (text), utterance identifier (utterance_id), data collection source (e.g., Accra or Kumasi), and duration (in seconds). Figure 1 provides a visual structure of the UGAkan-ImpairedSpeechData corpus.

Correspondence: aoekpezu@st.ug.edu.gh

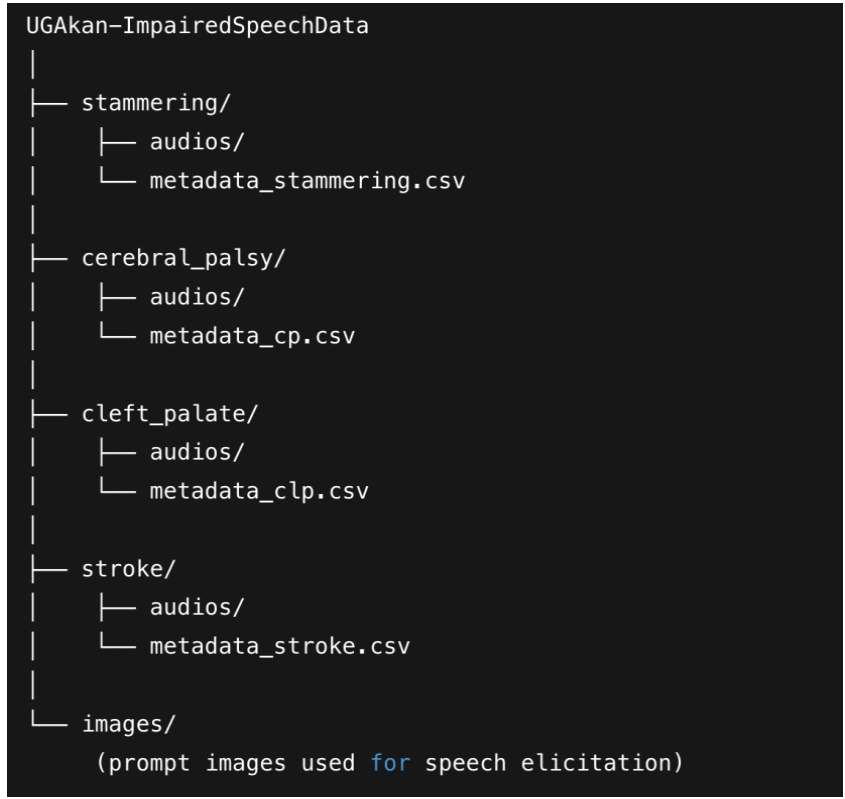

Figure 3. Visual structure of the impaired speech corpus

Table 1 summarizes the distribution of the dataset by aetiology, gender, number of audio files, and total duration in hours. Cerebral Palsy represents the largest subset (251 MB), followed by Stammering (198 MB), Cleft Palate (55.2 MB), and Stroke (7.53 MB). The average recording length is 12.46seconds, with most samples between 6.59seconds and 16seconds.

Table 5: Distribution of audio Files and total duration (in hours) by aetiology and gender

|  | Aetiology | Total no of audio files | Equivalent in hours |
|---|---|---|---|
| Male | Cerebral Palsy | 2,881 | 8.84 |
|  | Cleft | 1,677 | 4.93 |
|  | Stroke-induced speech disorder | 295 | 0.74 |
|  | Stammering | 1,901 | 5.51 |
| Female | Cerebral Palsy | 4,835 | 15.66 |
|  | Cleft | 149 | 0.45 |
|  | Post-Stroke | 0 | 0 |
|  | Stammering | 2,574 | 13.88 |
| Total number of unique files |  | 14,312 | 50.01 |

3.2  Value of the Data

i. The dataset addresses both a low-resource language (Akan) and a low-resource speech domain (impaired speech). Particularly, it offers the first publicly available impaired speech dataset in Akan. The combination of these underrepresented areas can enhance advancements in inclusive speech technologies in Ghana and neighbouring Akan speaking countries.

ii. The dataset supports research in automatic disordered speech recognition (ADSR). It will enable the adaptation of existing automatic speech recognition (ASR) models to impaired speech. This will be particularly useful for developing systems that can

Correspondence: aoekpezu@st.ug.edu.gh

      accurately recognize the speech patterns of individuals with speech impairments such as cerebral palsy, cleft palate, stammering, and stroke-induced speech disorders in Akan.
iii. The dataset enables the design of assistive communication tools for individuals with speech impairment. It can be used to develop assistive technologies that interpret or amplify the speech of individuals with impairments, thereby improving communication in clinical, educational, or everyday social settings.
iv. The dataset covers four distinct aetiologies of speech impairment. Thus, it supports diverse research applications. Researchers, clinicians, and technology developers can leverage the dataset to build and evaluate inclusive, culturally grounded AI models for speech recognition and synthesis in low-resource settings. With data from individuals affected by cerebral palsy, cleft palate, stammering, and stroke, the dataset enables comparative studies on speech patterns across different impairment types. It also supports research into multi-condition ASR model generalization. Researchers can reuse the data for training, benchmarking, or augmenting models in ASR, sociophonetics, and clinical linguistics.
v. The dataset includes 50.01 hours of audio recordings with corresponding transcriptions. The alignment of speech and text enables its use in language technology applications such as automatic speech recognition (ASR), text-to-speech (TTS) synthesis, and other natural language processing (NLP) tasks.

## 4 Limitations

A key limitation of this study is the use of convenience sampling due to social and logistical constraints in recruiting individuals with severe speech impairments. This restricted our ability to include a balanced and diverse representation across all severity levels and impairment types. It was difficult to recruit participants from certain groups, such as individuals with cleft palate conditions and especially those recovering from stroke. Stroke cases were the most underrepresented, with only 296 entries collected, all from a specific gender. In cases involving motor impairments like cerebral palsy, scalable data collection methods such as unsupervised or remote recording were not feasible. Ideally, recording individuals with severe impairments in a controlled studio setting would help better capture their speech patterns. However, this was not practical, as it would have significantly extended the data collection period. Additionally, the dataset does not audio recordings with extreme severe impairments. This further narrows the range of impairment severity represented. These limitations may introduce sampling bias and reduce the dataset's generalizability, particularly for developing models tailored to the most severe and underrepresented speech impairment cases.

## 5 Ethics Statement

Ethical approval for this study was obtained from the Ethics Committee for Basic and Applied Sciences (ECBAS) at the University of Ghana. All participants were fully informed about the objectives of the audio data collection and the potential benefits of developing a speech dataset. Although personal information such as names and phone numbers was collected, it was used exclusively for administrative purposes and to facilitate compensation.

| **Data accessibility** | Repository name: UGAkan-ImpairedSpeechData: A Dataset of Impaired Speech in the Akan Language<br>Data identification number: 10.17632/vc84vdw8tb.3 |
|---|---|

Correspondence: aoekpezu@st.ug.edu.gh

| | Direct URL to data: https://data.mendeley.com/datasets/vc84vdw8tb/4 |
|---|---|
| **Related research article** | One half of the dataset was utilized for evaluating and adapting ASR models [4]. |

# 6 Acknowledgements

This research was funded by Google Research Ghana. We thank all the individuals who contributed to the success of this project but do not meet the criteria for authorship. Special thanks go to the speech and language therapists, language experts, and individuals who provided technical support who dedicated their time and effort to ensure the quality and completeness of the dataset. We are also grateful to the community leaders, institutions, and caregivers who supported participant recruitment and provided logistical assistance during data collection.

Correspondence: aoekpezu@st.ug.edu.gh